\documentclass[aps,prb,manuscript]{revtex4}
\usepackage[centertags]{amsmath}
\usepackage{amssymb}
\usepackage{amsthm}
\usepackage{graphicx}
\usepackage{epsfig}
\draft
\begin{document}

\title{Noise correlation-induced splitting of Kramers' escape rate from a metastable state}

\author{Pulak Kumar Ghosh$^a$,  Bidhan Chandra Bag$^b$
and Deb Shankar Ray$^a${\footnote {e-mail address:
pcdsr@mahendra.iacs.res.in}}}

\affiliation{$^a$Indian Association for the Cultivation of Science,
Jadavpur, Kolkata
700 032, India\\
$^b$Department of Chemistry, Visva-Bharati, Santiniketan 731 235,
India}

\noindent
\begin{abstract}
A correlation between two noise processes driving the thermally
activated particles in a symmetric triple well potential, may cause
a symmetry breaking and a difference in relative stability of the
two side wells with respect to the middle one. This leads to an
asymmetric localization of population and splitting of Kramers' rate
of escape from the middle well, ensuring a preferential distribution
of the products in the course of a parallel reaction.
\end{abstract}

 \maketitle
\section{Introduction}
The escape of particles from a metastable well has long been the
focal theme of activated rate
processes\cite{kram,hangi,ff22,san,db11} in chemical kinetics and
condensed matter physics. When a stochastic system is
thermodynamically made open by the action of external periodic or
random forces significant changes in the dynamics take place which
reflect the constructive role of noise in dynamical systems. The
well-known examples include stochastic
resonance\cite{ben,luc,mcn,d41}, resonant
activation\cite{dor,Bier,van,pk1}, noise-induced
transition\cite{sch,hor,fuli}, ratchet and rectification of
noise\cite{ajd,mag1,kul,pkg2,march12,pcbcb} etc to mention a few. In
the overwhelming majority of these cases the essential physics
relies on double-well as a model potential. The present work
concentrates on the motion of Brownian particles in the middle well
of a symmetric triple well potential which diffuses symmetrically to
the left and the right well. At a finite temperature and in absence
of any bias force the particles are activated only by inherent
thermal fluctuations resulting in equalization of population in the
two side wells. Application of an additive noise can not lead to any
change of time averaged relative population of the two wells. The
situation is expected to remain unaltered even if, in addition, one
introduces a multiplicative noise which makes the diffusion state
dependent. The aim of the present work is to look for a scheme which
leads to a symmetry breaking resulting in a preferential population
distribution in one of the two side wells. In what follows we show
that the presence of correlation between the applied additive and
multiplicative noises may cause a change in relative stability of
the side wells with respect to the middle one. This correlation
induced interference of the two noises leads to a splitting of
Kramers' rate of escape from the metastable well.

The correlation between noise processes has been the subject of
study in a number of issues. For example, it has been shown that
correlation strongly influences the noise-induced phase transitions
from unimodal to bimodal distribution \cite{fuli}. Fox\cite{fox1}
has investigated the correlation between multi-component, Markovian
and Gaussian stochastic processes. The effect of correlation between
quantum noises\cite{singh} in laser modes and in the description of
hydrodynamic modes\cite{fedc} are of interest in the related
context. Our proposal in this work concentrates on altering the
relative stability of the two side wells with respect to the middle
one of a triple-well potential under the influence of correlation
between the noises, where the underlying idea rests on controlling
the pathways of a parallel reaction. A prototypical example may be
set by considering a nucleophilic attack by $X^{-}$ (a halid ion of
$HX$) at the carboxyl group of a ketone say $R_1(R_2)C=0$ which
produces $D$ and $L$ forms of $R_1(R_2)C(OH)X$ - the two optical
isomers. They are of same energy but differ in optical properties.
The middle well represents the reactant state while the two side
wells refer to the product states of the parallel reaction. To
realize an experimental situation we introduce a light field of
fluctuating intensity which polarizes the photosensitive carbonyl
group rendering planarity of the polarized states and causing
symmetric oscillation of the barrier height. If an electric field of
fluctuating intensity and of common origin is now imposed, in
addition, on the polarized system, then depending on the
cross-correlation of the fluctuations of light and electric fields,
the nucleophilic attack of an anion $X^-$ will be asymmetric. This
is because the electric field rocks the side wells (similar to what
one observes in stochastic resonance) implying the differential
relative stability of the transition states of the complex
comprising polarized system plus anion. Based on a Langevin and an
associated Fokker-Planck description for the dynamics where the
additive and multiplicative noise processes are independently
Gaussian and $\delta-$correlated in character but the
cross-correlation between them is exponential, we derive an
analytical expression for the splitting of Kramers' rate due to
correlation induced asymmetry in the state dependent diffusion. This
results in preferential distribution of reaction products of the
parallel reaction due to differential relative stability of the two
wells.
\section{The model}
Consider an overdamped Brownian particle in a triple-well potential
$V(x)$ kept in a thermal bath at temperature $T$ and subjected to
two stochastic forces $\epsilon_1(t)$ and $\epsilon_2(t)$. The
governing Langevin equation is given by
\begin{eqnarray}\label{1}
\gamma\dot{x}= -V'(x)\;+\;\epsilon_1(t)+\;x\epsilon_2(t)+\Gamma(t)
\end{eqnarray}
where $V(x)=x^2(bx^2-c)^2$ symmetric triple-well potential (Fig.1);
$b$ and $c$ are the parameters of the potential and $\gamma$ is the
dissipation constant.  Thermal fluctuation $\Gamma(t)$ of the bath
is modeled by Gaussian, zero mean  and delta correlated noise
\begin{subequations}
\begin{eqnarray}
\langle \Gamma(t)\rangle&=&0\\\label{2a}
 \langle
\Gamma(t)\Gamma(t')\rangle&=&2 D \delta(t-t')\label{2b}
\end{eqnarray}
\end{subequations}
$D$ being the strength of thermal fluctuation and is given by
$D=kT/\gamma$. Here the additive random force $\epsilon_1(t)$ rocks
the potential wells sidewise randomly, whereas the multiplicative
force $\epsilon_2(t)$ sets random fluctuation of the barrier height
around $\Delta V_0( = {4c^3}/{27b})$, in a symmetric manner. The
system as described by (\ref{1}) is associated with both the thermal
and non-thermal environments. To achieve our desired asymmetry in
the dynamics of the particle in a symmetric triple-well potential,
we apply an electric field and a radiation field simultaneously. The
interaction with the radiation field is relatively stronger than
that with the applied electric field. The multiplicative and
additive noises in Eq.(\ref{1}) correspond to fluctuating amplitude
of the radiation field and the electric field, respectively. To keep
the treatment on a general footing one may assume, $\epsilon_1$ and
$\epsilon_2$ to be colored. This may cause a serious difficulty for
an analytical approach. On the other hand to capture the essential
physics we may assume $\epsilon_1(t)$ and $\epsilon_2(t)$ to be
Gaussian white noises. This does not change the inherent feature of
the proposed model problem. The characteristics of the noise
processes can be summarized as follows
\begin{subequations}
\begin{eqnarray}
\langle \epsilon_1(t)\rangle&=&\langle \epsilon_2(t)\rangle
=0\\\label{3a} \langle \epsilon_1(t)\epsilon_1(t')\rangle&=& 2\;
Q_1 \delta(t-t')\\\label{3b} \langle
\epsilon_2(t)\epsilon_2(t')\rangle&=& 2 \;Q_2
\delta(t-t')\label{3c}
\end{eqnarray}
\end{subequations}
$Q_1$, $Q_2$ are the strength of $\epsilon_1(t)$ and
$\epsilon_2(t)$. Now if the simultaneous action of fluctuating
electric and radiation fields, is due to a common origin then the
statistical properties of the noises are not expected to differ
widely and may be correlated \cite{li,mad,jia1}. We characterize the
correlation of $\zeta(t)$ and $\eta(t)$  as
follows\cite{xia,jia,jia1,mei,mei1,pb}
\begin{eqnarray}
\langle \epsilon_1(t)\epsilon_2(t')\rangle &=& \langle
\epsilon_1(t')\epsilon_2(t)\rangle = \frac{\lambda\sqrt{Q_1
Q_2}}{\tau}\;\exp{\left[- \frac{(t-t')}{\tau}\right]}\\\label{3d}
\;\;&=&2\;\lambda\sqrt{Q_1 Q_2}\;\delta(t-t')  \;\;\;\;\;\;\;\;
as \; \;\tau \rightarrow 0\nonumber
\end{eqnarray}
$\lambda$ is strength of cross-correlation and $\tau$ is the
cross-correlation time.  By colored correlation between the white
noises we mean that both the external fluctuations are affected by
each other for certain ranges frequencies. Again if we assume that
cross-correlation is a $\delta-$ function, then the gross feature of
our model problem will remain unchanged. However for generality, we
have assumed colored cross-correlation.

We now proceed with a probabilistic description corresponding to
Langevin equation(\ref{1}) with the prescriptions(2,3,4) for
internal thermal noise and external forces, respectively.
Following\cite{san,san1,str,jai,xia}, the time evolution equation
for the probability density is given by
\begin{eqnarray}
\frac{\partial P(x,t)}{\partial t}=-\frac{\partial}{\partial
x}V'(x)P(x,t)-\frac{\partial}{\partial x}\langle \epsilon_1(t)
\;\delta(x(t)-x)\rangle &-&\langle x \epsilon_2(t)
\;\delta(x(t)-x)\rangle\nonumber\\
 &+& D\frac{\partial ^2 P(x,t)}{\partial
x^2}\label{4}
\end{eqnarray}
\noindent where $P(x, t)=\langle \delta(x(t)-x)\rangle$; the
averages $\langle...\rangle$ in Eq.(\ref{4}) can be calculated for
Gaussian noise by the Novikov theorem\cite{nov1}. The resulting
equation is the Fokker-Planck description as given by\cite{jai,xia}.
\begin{eqnarray}
\frac{\partial P(x,t)}{\partial t} =-&&\frac{\partial}{\partial x}
f(x) P(x,t) +Q_2\frac{\partial}{\partial x}x\frac{\partial}{\partial
x}x P(x,t)+\frac{\lambda\sqrt{Q_1 Q_2}}{1+8c^2\tau}
\;\frac{\partial}{\partial x}x\frac{\partial}{\partial x}
P(x,t)\nonumber
\\&&+ \frac{\lambda\sqrt{Q_1 Q_2}}{1+8c^2\tau}
\;\frac{\partial^2}{\partial x^2}x
P(x,t)\;+(Q_1+D)\frac{\partial^2}{\partial x^2}P(x,t) \label{5}
\end{eqnarray}
The only constraint on $\tau$ is that,
\begin{eqnarray}
 1+8c^2\tau > 0\label{6}
\end{eqnarray}
$c$ is a potential parameter and a real positive number. Thus
practically there exists no restriction on $\tau$ in this case.

\section{Stationary distribution and asymmetric localization}
We now return to the Fokker-Planck Eq.(\ref{5}). We can recast it in
the more simpler form as follows
\begin{eqnarray}
\frac{\partial P(x,t)}{\partial t} =-\frac{\partial}{\partial
x}\left[ f(x) -\frac{1}{2}\frac{\partial}{\partial
x}D(x,\tau)-D(x,\tau)\frac{\partial }{\partial x}
\right]P(x,t)\label{8}
\end{eqnarray}
where $D(x,\tau)$ is the effective diffusion constant
\begin{eqnarray}
D(x,\tau) = Q_2 x^2+2\left(\frac{\lambda\sqrt{Q_1 Q_2}}{1+8c^2\tau}
\right) x +Q_1+D \label{9a}
\end{eqnarray}
The effective diffusion constant is an asymmetric function of
position, which means it is dissimilar in the three wells of the
triple well potential. Around the minimum of the middle well
diffusion is almost independent of space and has a constant value,
while towards the right well diffusion strength quadratically
increases with position and decreases towards the left well. So the
implications for the  simultaneous action of the two forces
$\epsilon_1(t)$ and $\epsilon_2(t)$ is that, the diffusive nature of
the Brownian motion in the three wells differs and therefore one can
expect an asymmetric distribution of the particles in the two side
wells and different transition rates from middle well to right and
left well.

As $t\rightarrow \infty$, the system reaches the stationary state
($\frac{\partial P}{\partial t}=0$) with the current attaining
 a constant or zero value. As the process $x(t)$ is bound
 to the triple-well potential it is expected  that in the stationary
 state there will have no net flow of particles, and hence
 one may assume a zero current stationary state.
 The solution of the Eq.(\ref{8}) is given by
\begin{eqnarray}
P(x)= D(x,\tau)^{-1/2}\exp{\left[\int^x dy
\frac{f(y)}{D(y,\tau)}\right]}\label{10}
\end{eqnarray}
It is apparent from the expressions (\ref{9a}) and (\ref{10}) that
as a result of interplay of two stochastic driving forces the
distribution function is asymmetric in space. The asymmetry in the
distribution function arises due to asymmetric diffusion of the
particles. To illustrate the asymmetric localization of the
particles we have plotted the distribution function as a function of
position in Fig.2. The solid line of Fig.2 presents a symmetric
distribution in absence the external driving forces. The dotted line
presents the same plot in presence of additive and multiplicative
noise but for no cross-correlation ($\lambda = 0$) . In this case
the distribution function still remains symmetric but the population
of the middle well is much higher than that of the side wells due to
the fact that in presence of multiplicative noise the effective
diffusion becomes space dependent, $D (x)=Q_2x^2+Q_1+D$.
 As a result the particles in the side wells diffuse to
the middle more quickly as compared to the particles of the middle
well which diffuse to the terminal wells at a relatively slower rate
and the particles spend most of the time in the middle well. This
implies that higher diffusion destabilizes the side wells compared
to the middle well. When the two stochastic forces are correlated
($\lambda \neq 0$), interestingly, because of the spatial asymmetry
in diffusion the Brownian particles are preferentially localized in
the left well compared to the right. This has been presented by the
dashed line in Fig.2.

To proceed further we require a quantifier which measures the
asymmetry in localization in the two wells. To this end we choose
the mean position of the particle as its measure. For a symmetric
distribution mean position $\langle x\rangle=0$  and for the
localization of the particles in left or right well, the value of
mean position is negative or positive, respectively. To this end an
expression for $\langle x \rangle$ from a direct steady state
solution of Fokker-Planck equation (10) can be formally obtained.
However, since this involves a complicated form of space dependent
diffusion coefficient, the final expression, which, in principle,
contains all the information regarding the interwell transitions, is
a lumped expression and it is difficult to figure out the details of
transition from one well to another. To have a closer look into this
aspect we examine the variation of mean position $\langle x \rangle$
with temperature, $Q_1$, $Q_2$ and correlation time with the help of
a discrete three-state model for the continuous triple well
potential. Three states are denoted by $x_0$, $\pm x_m$ for the
symmetric unperturbed system corresponding to three minima. The
diffusional motion causes transitions between them and it is
schematically presented by the following kinetic model
\begin{eqnarray}
\;\;\;\;k_{L}\;\;\;\;\;\;\;\;\;\;\;&k_{M}^R& \nonumber\\
L\;\;\;\;  \rightleftharpoons\;\;\; M \;\;\;&\rightleftharpoons& \;\;\;\;R \nonumber\\
\;\;\;\; k_{M}^L \;\;\;\;\;\;\;\;\;\;\;&k_{R}& \nonumber
\end{eqnarray}
$k_L,\;k_R,\; k_M^R ,\; k_M^L $ denote the time averaged rate of
transition from left to middle well, right to middle well, middle to
right and middle to left well, respectively. The  number of
particles in  the three states at time $t$
 are denoted by $n_{L},\; n_R$ and $n_L$. The governing master equations for
 $n_{i}$ ($i=L,\; R,\; M$) read as
 \begin{eqnarray}
\frac{dn_L}{dt}&=&-k_L\;n_L+ k_M^L \;n_M\\\label{11}
\frac{dn_R}{dt}&=&-k_R\;n_R+ k_M^R \;n_M\\\label{12}
\frac{dn_M}{dt}&=&k_L\;n_L+k_R\;n_R -( k_M^L +k_M^R )\;n_M
\label{13}
\end{eqnarray}
At the steady state ($\dot{n_L}=\dot{n_M}=\dot{n_R}=0$) the
probability of finding the particles at the three wells $P_{i}$
($i=L,\; R,\; M$) are
\begin{eqnarray}
P_L= k_M^L k_R/P,\;\;\;\;\;\;\; P_R=k_M^R  k_L/P,\;\;\;\;\; \;
P_M=k_L k_R/P \label{14}
\end{eqnarray}
 where $P=k_R  k_M^L +k_R k_L+k_L  k_M^R$.
The expression for the mean position is then given by
\begin{eqnarray}\label{9}
\langle x \rangle &=&\int_{-\infty}^{+\infty}x P(x)\;dx=x_m
P_R+x_0P_M-x_mP_L\nonumber
\\
&=&\left(\frac{\sqrt{27\Delta V_0}}{2c}\right)\frac{ k_M^R
k_L-k_M^L  k_R}{k_R k_M^L +k_R k_L+k_L k_M^R}\label{15}
\end{eqnarray}
The above expression clearly expresses the dependence of mean
position and probability on four rate constants. All the individual
rate constants have a typical dependence on the system parameters,
such as, temperature, intensity of the external noises and
cross-correlation time. We therefore anticipate a distinct signature
of asymmetric localization of the particles with the variation of
system parameters. As revealed by Eq.(\ref{15}) the mean position is
directly proportional to ($k_M^R k_L-k_M^L  k_R$), that is, the
difference involving the product of right hand directed transition
rates and the product of left hand directed transition rates. Our
numerical illustration shows that the mean position $\langle x
\rangle \neq 0$ only if $\lambda\neq 0$. This means that the
particles are asymmetrically localized in the triple-well potential
only when there is a finite correlation between two external
stochastic drives $\epsilon_1(t)$ and $\epsilon_2(t)$. This type of
behaviour can be physically explained as follows: As both the
external forces $\epsilon_1(t)$ and $\epsilon_2(t)$, act
independently to rock the potential well randomly in an asymmetric
manner and randomly modulate the barrier heights in a symmetric way,
respectively, then the symmetry of the triple-well system remains
intact as the individual action of $\epsilon_1(t)$ and
$\epsilon_2(t)$ are not able to break the symmetry of the system. If
there is some correlation between $\epsilon_1(t)$ and
$\epsilon_2(t)$, which means that for a number of stochastic
realizations, both forces have the same sign at a particular instant
of time. A qualitative interpretation of the localization may be
given as follows: as long as the force $\epsilon_2(t)$ causing
symmetric fluctuation of the barrier height attains its lower value,
the force $\epsilon_1(t)$ points to the right well so that the
particle in the middle well move towards the right well very
quickly. On the other hand as the tilting force points to the left
$\epsilon_2(t)$ sets the barrier height at a larger value and
consequently the particle in the middle well takes relatively larger
time to speed up from middle to left well for the simultaneous
action of the synchronized forces. The particles in the middle well
therefore have a greater chance to cross the right-hand barrier and
the particles in the right well have a greater chance to move to the
middle well as the fluctuation of barrier height in the right well
occurs with a higher amplitude. Therefore the relative stability of
the two wells with respect to the middle differs and an asymmetric
localization is observed.

Another important question on the asymmetric localization concerns
in which well the particles will be preferentially localized, out of
two side wells of the triple-well potential. Secondly, what will be
the sign of $\langle x \rangle $. An answer to this question may be
obtained as follows. As the diffusive motion of the particles in the
right well is greater than that in the left well (as given by the
Eq.(\ref{9})), the particles will prefer to be localized in the left
well, since higher diffusion makes the right well relatively less
stable. So the sign of $\langle x \rangle $ will be always negative.
The sign of $\langle x \rangle $ and hence asymmetric localization
can be inverted by reversing the sign of $\epsilon_1(t)$ in the
Eq.(\ref{1}).

We now proceed to analyze the behavior of localization (mean
position) with the variation of system parameters. The effect of
temperature in asymmetric localization and in the product
distribution of the parallel reaction at the steady state is
intimately related to the manipulation of inherent condition rather
than coherence in selecting and controlling the reaction pathways.
Keeping in view of the Arrhenius temperature dependence of the
individual rate constants, the variation of $\langle x \rangle$ with
temperature is expected to show a bell-shaped curve. The departure
of $\langle x \rangle$ from zero towards negative direction
indicates the preferential distribution of the product in the left
well. The variation of mean position $\langle x \rangle$ as a
function of temperature for several values of strength of
correlation for the input forces as shown in Fig.3 corroborates this
assertion. Fig.3 also reveals that for fixed values of temperature
and other parameter set the mean position increases with increase of
the strength of cross-correlation. It clearly indicates that, for an
appropriate correlation between additive and multiplicative noises,
the particles will be preferentially localized more asymmetrically.
In Fig.4 we show the mean position as a function of intensity of the
multiplicative noise.  With the increase of values of $Q_2$ the mean
position gradually moves to a maximum negative value followed by a
return to zero at high intensity ($Q_2$). This sort of behavior can
be understood from the expression for the effective diffusion
constant(\ref{9a}). In this expression $Q_2$ appears as the
symmetric contribution  in the first term and as the asymmetric
contribution (as $\sqrt{Q_2}$) in the second term, so that if one
starts from the very low value of $Q_2$, the asymmetry in diffusion
first increases then at a relatively high value it starts
decreasing. In Fig.5 we present the variation of mean position as a
function of intensity of the additive driving force. With increase
of $Q_1$ the value of mean position starts departing from zero to
reach finally a limiting value. Finally, in Fig.6 we have examined
the effect of the cross-correlation time in the asymmetric
localization by plotting the mean position as a function of $\tau$
for different values of the strength of cross-correlation. As the
asymmetry in diffusion decreases with increase of $\tau$, the
departure of the mean position decreases with increasing values of
cross-correlation time.

\section{Transition rate from middle well to side wells:
Splitting of Kramers' rate}

To what extent the correlation induced asymmetry in the effective
diffusion coefficient is reflected in the kinetics of activated
processes. An answer to this question lies in examining the
transition rates of the particles from the middle well to the side
wells. The differential behavior of the transition rates signifies
the scope of controlling the path of a parallel reaction. We
approach the problem by calculating the mean escape time. The
expression for MPFT \cite{str} (mean first passage time) for a
particle to reach the final point $\pm x_b$, starting from an
initial point $x_0$ is given by
\begin{subequations}
\begin{eqnarray}
T_R&=&\int^{+x_b}_{x_0} \frac{dx}{D(x,\tau)P(x)}\int_{-\infty}^x
P(y)dy\label{16a}
\\
T_L &=&\int^{-x_b}_{x_0} \frac{dx}{D(x,\tau)P(x)}\int_{-\infty}^x
P(y)dy\label{16b}
\end{eqnarray}
\end{subequations}
respectively, where $T_R$ and $T_L$ denote the mean escape time of
the particle from $x_0$ to $+x_b$ and $-x_b$ respectively ($+x_b$
and $-x_b$ are the coordinates of barrier tops toward right and left
well, respectively). Putting the expressions for $P(x)$ and $D(x)$,
the integrals in the above Eqs(\ref{16a},\ref{16b}) have been
calculated using steepest-descent approximation to obtain the
expressions for $T_i$ ($i=R, \;L$) in the usual way
\begin{subequations}
\begin{eqnarray}
T_R&=&\frac{2\pi}{\sqrt{\omega_{R}
\omega_0}}\exp{\left[-\int_{x_0}^{+x_m}
dx\;\frac{f(x)}{D(x,\tau)}\right]}\label{17a}
\\
T_L&=&\frac{2\pi}{\sqrt{\omega_{L}
\omega_0}}\exp{\left[-\int_{x_0}^{-x_m}
dx\;\frac{f(x)}{D(x,\tau)}\right]}\label{17b}
\end{eqnarray}
\end{subequations}
where $\omega_0,\;\omega_R,\; \omega_L$  are the frequencies
corresponding to the potential minimum ($x_0$) and the barrier tops
($\pm x_b$), respectively. As $D(x,\tau)$ is an asymmetric function,
the integrals $\int_{x_0}^{+x_b} dx\;\frac{f(x)}{D(x,\tau)}$ and
$\int_{x_0}^{-x_b} dx\;\frac{f(x)}{D(x,\tau)}$ are not same. It is
thus apparent from the above expression that the transition rate
from the middle well to the terminal wells splits up due to the
interplay of two correlated stochastic forces. The ratio of the
transition rates ($k_M^R/k_M^L$) deviates from unity
($T_L/T_R=k_M^R/k_M^L\neq 1$) only when the two external drives are
correlated ($\lambda\neq 0$). If $\lambda=0$, $D(x,\tau)$ is a
symmetric function of $x$ and $\int_{x_0}^{+x_b}
dx\;\frac{f(x)}{D(x,\tau)}=\int_{x_0}^{-x_b}
dx\;\frac{f(x)}{D(x,\tau)}$. So the ratio of the transition rates
become unity. In Fig.7 we present the variation of the ratio of the
transition rates($k_M^R/k_M^L$) as a function of temperature for
several values of the coupling strength. As revealed by Fig.7, the
ratio of the transition rates at a very low temperature
significantly differs from unity and tends to equalize in the high
temperature limit. For a fixed temperature the ratio of the
transition rates deviates more from unity for an increase of the
coupling strength($\lambda$). In Fig.8(a,b) we plot the ratio of the
transition rates as a function of the intensities ($Q_1, Q_2$) of
the external noises. One observes that with increase of intensities
of the external noises the magnitude of $k_M^R/k_M^L$ increases to
maximum followed by a decrease. This resonance like behavior is
observed due to the fact that with increase of intensities of the
external noise the synchronization probability of the noise
realizations from two noise processes( $\epsilon_1(t),
\;\epsilon_2(t) $) in a given time increases as a result of which
the transition probability to the right increases. Further increase
of noise intensities results in randomization of the system. In
order to examine the influence of cross-correlation time of the
nonthermal noises on the the ratio of transition rate we plot in
Fig.9 the variation of $k_M^R/k_M^L$ as a function of $\tau$ for
several values of $\lambda$. With increase of cross-correlation time
the ratio of the transition rates monotonically decreases. This is
again due to the asymmetry in the diffusion coefficient which
decreases with increase of $\tau$.

\section{conclusion} We have considered the stochastic dynamics of the
particles in a triple well potential driven simultaneously by two
cross-correlated white noise processes. It has been shown that
depending on the correlation between the two noise sources, one
multiplicative and another additive, the relative stability of the
two side wells with respect to the middle one may differ
significantly. This originates from an asymmetry in diffusive motion
of the interwell dynamics due to interference of the two noises. An
offshoot of this symmetry breaking effect is the splitting of
Kramers' rate of escape from the middle well to the sides wells. In
a wider context the kinetic scheme may serve as a technique for
preferentially selecting a pathway for a parallel chemical reaction.
\acknowledgments Thanks are due to the Council of Scientific and
industrial research, Govt. of India, for partial financial support.

\newpage
\begin{center}
{\bf Figure Captions}
\end{center}

Fig.1 A schematic illustration of a symmetric triple-well potential
as an energy profile of the parallel reaction (specifically, optical
isomerization reaction).
\newline Fig.2. Probability distribution function $P(x)$ vs position $x$
plot depicting the
 changes of distribution due to the addition of external noises for the
 parameter set: $\tau=0,\;T=0.44, \;b=0.1 $ and $c=1.0$. (i)
 Solid line presents the distribution function in absence of external
 stochastic forces, (ii) dotted line presents the same plot in
 presence of external additive and multiplicative noise
 in absence of cross-correlation, (iii) dashed line presents also same
 plot in presence of external additive and multiplicative noise and
 their cross-correlation.

Fig.3. Variation of mean position $ \langle x \rangle $ as a
function of temperature $T$ for several values of coupling strength
and for the parameter set: $\tau=0.01,\; Q_1=0.02, \;
Q_2=0.02,\;b=0.1 $ and $c=1.0$.

 Fig.4. Mean position($\langle x \rangle$) vs $Q_2$ (strength of
the multiplicative noise) plot for several values of coupling
strength and for the parameter set: $\tau=0.01,\; Q_1=0.02, \;
T=0.34,\;b=0.1 $ and $c=1.0$.

Fig.5. Mean position($\langle x \rangle$) vs $Q_1$ (strength of the
multiplicative noise) plot for several values of coupling strength
and for the parameter set: $\tau=0.01,\; Q_2=0.02, \; T=0.34,\;b=0.1
$ and $c=1.0$.

Fig.6. Variation of $ \langle x \rangle $ as function of
cross-correlation time $\tau$ for different values of coupling
strength and for the parameter set: $T=0.34, Q_1=0.02,
Q_2=0.02,b=0.1 ,c=1.0$.

Fig.7. Ratio of transition rates ($k_M^R/k_M^L$) vs temperature plot
for several values of coupling strength and for the parameter set:
$\tau=1.0,\; Q_2=0.05, \; Q_1=0.05,\;b=0.1 $ and $c=1.0$.

 Fig.8. (a) Variation of the ratio of the transition rate
($k_M^R/k_M^L$) as function of the strength of additive noise and
for several values of coupling strength and for the parameter set:
$T=0.05,\; \tau=1.0, \; Q_2=0.02,\;b=0.1 $ and $c=1.0$. (b)
Variation of the ratio of the transition rate ($k_M^R/k_M^L$) as
function of the strength of multiplicative noise for same parameter
set as (a) but for $Q_1=0.05$.

Fig.9. Variation of the ratio of the transition rate ($k_M^R/k_M^L$)
as function of cross-correlation time $\tau$ for different values of
coupling strength and for the parameter set: $T=0.05,\; Q_1=0.05, \;
Q_2=0.05,\;b=0.1 $ and $c=1.0$.

\begin{figure*}[tp]
\centering
\includegraphics*[width=15.2cm]{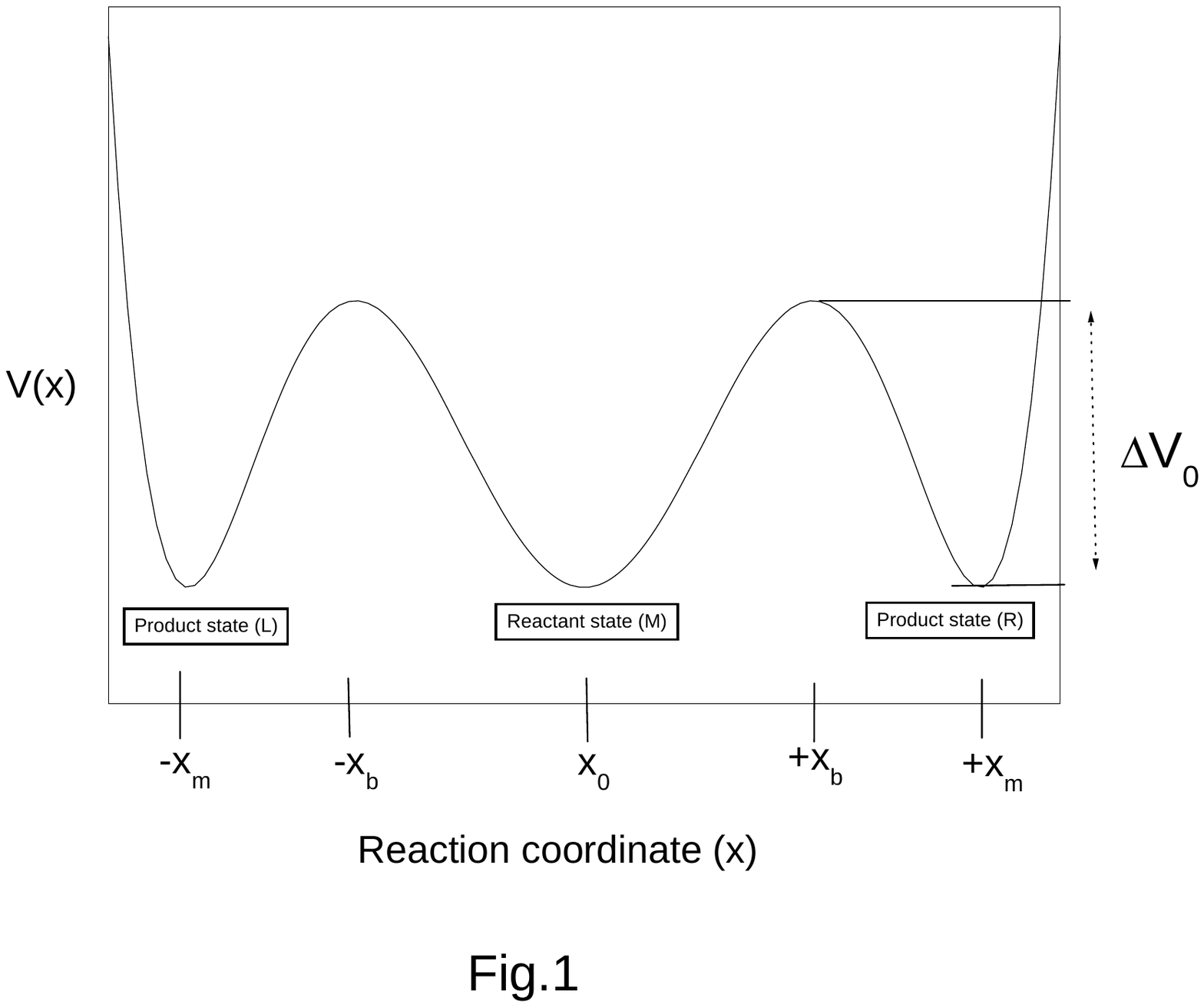}
\end{figure*}

\begin{figure*}[tp]
\centering
\includegraphics*[width=20.2cm]{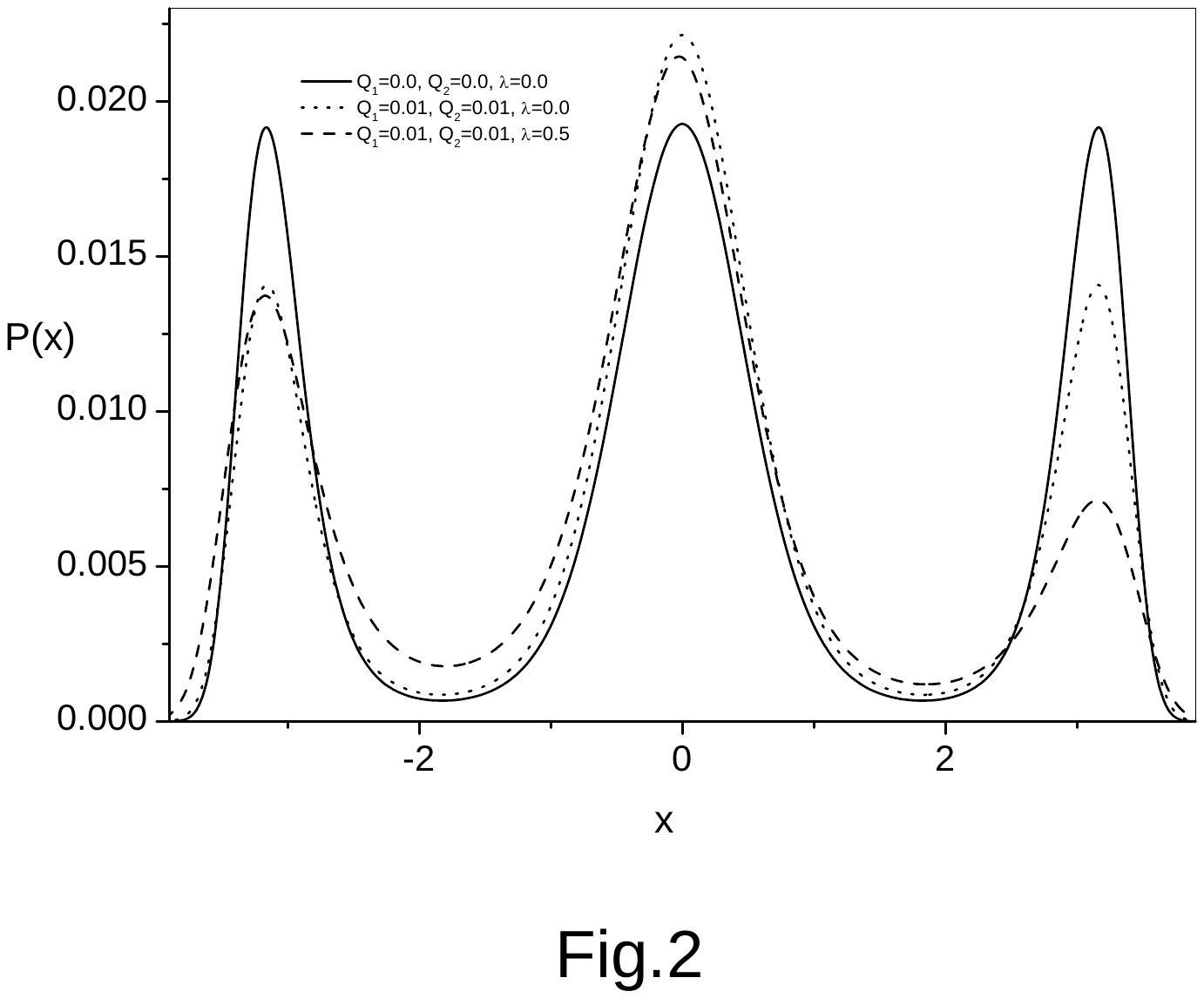}
\end{figure*}

\begin{figure*}[tp]
\centering
\includegraphics*[width=20.2cm]{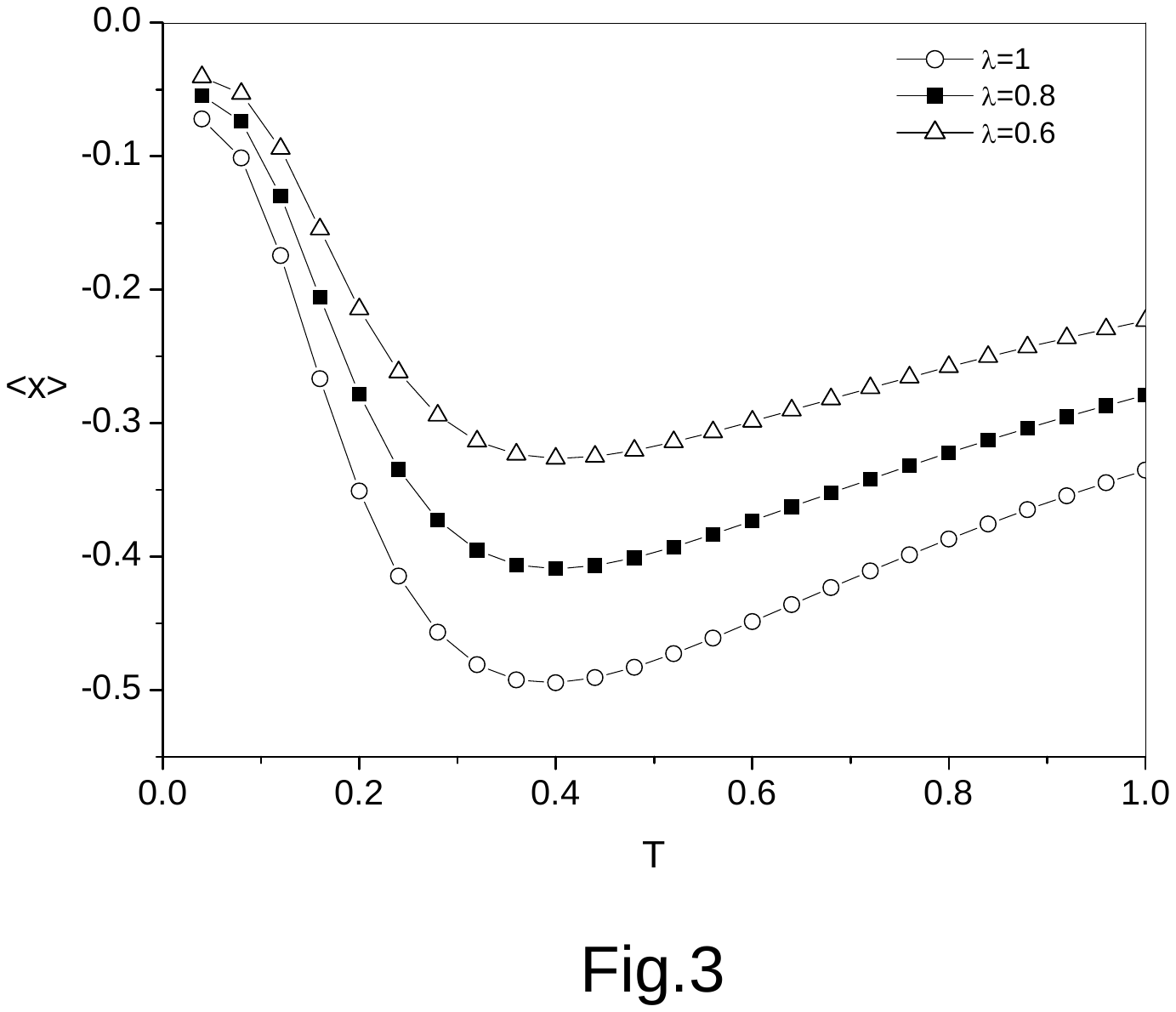}
\end{figure*}

\begin{figure*}[tp]
\centering
\includegraphics*[width=20.2cm]{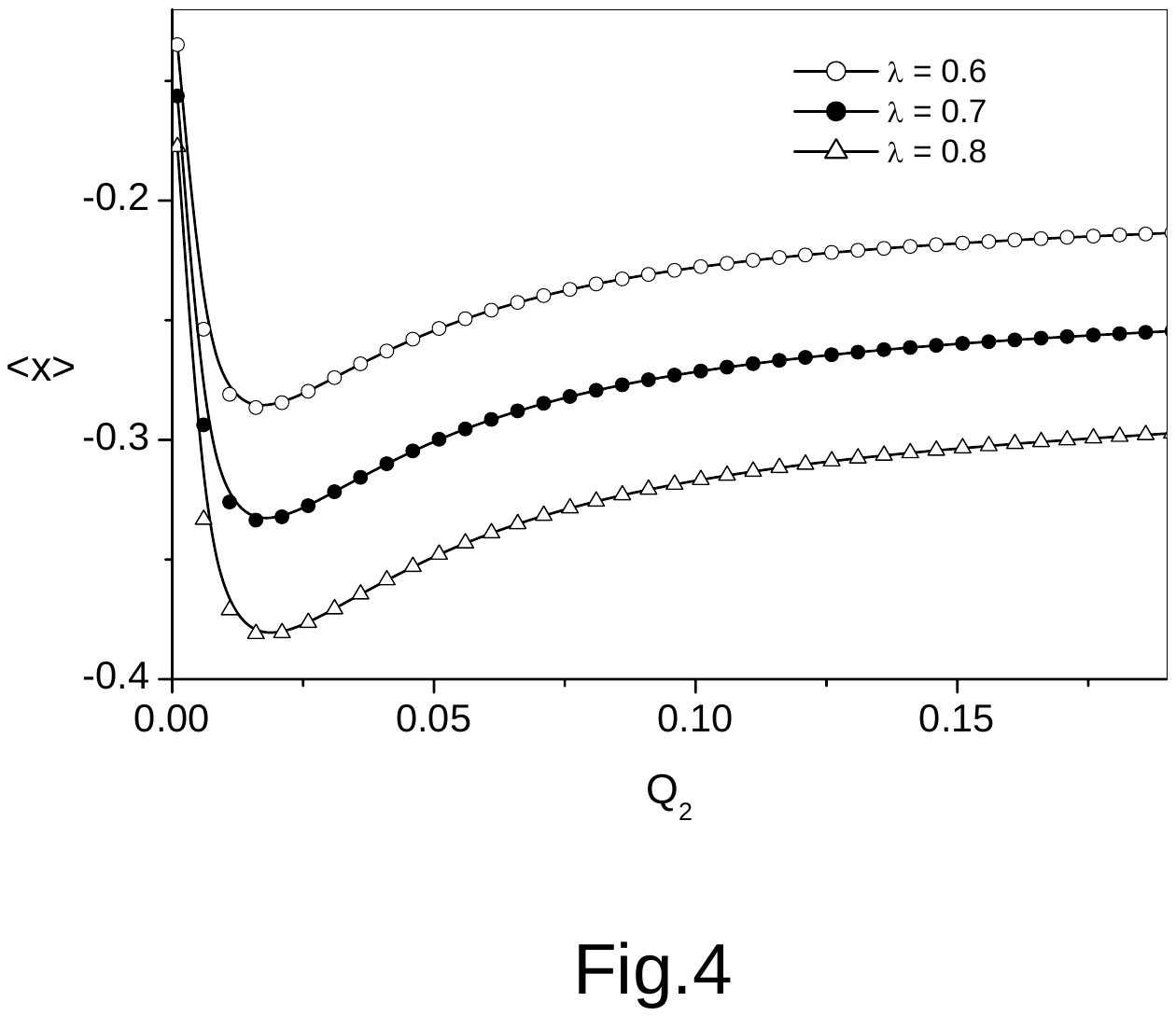}
\end{figure*}

\begin{figure*}[tp]
\centering
\includegraphics*[width=20.2cm]{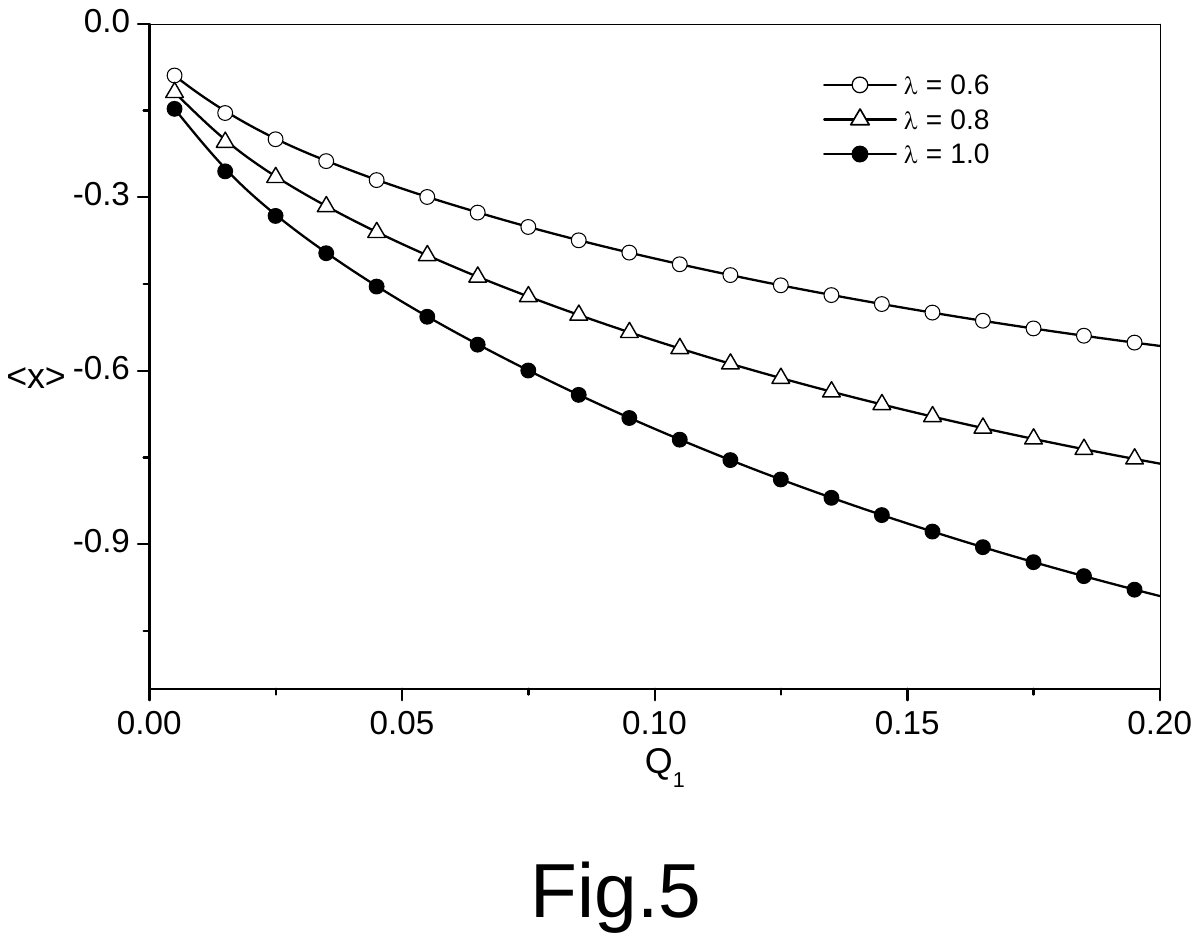}
\end{figure*}

\begin{figure*}[tp]
\centering
\includegraphics*[width=20.2cm]{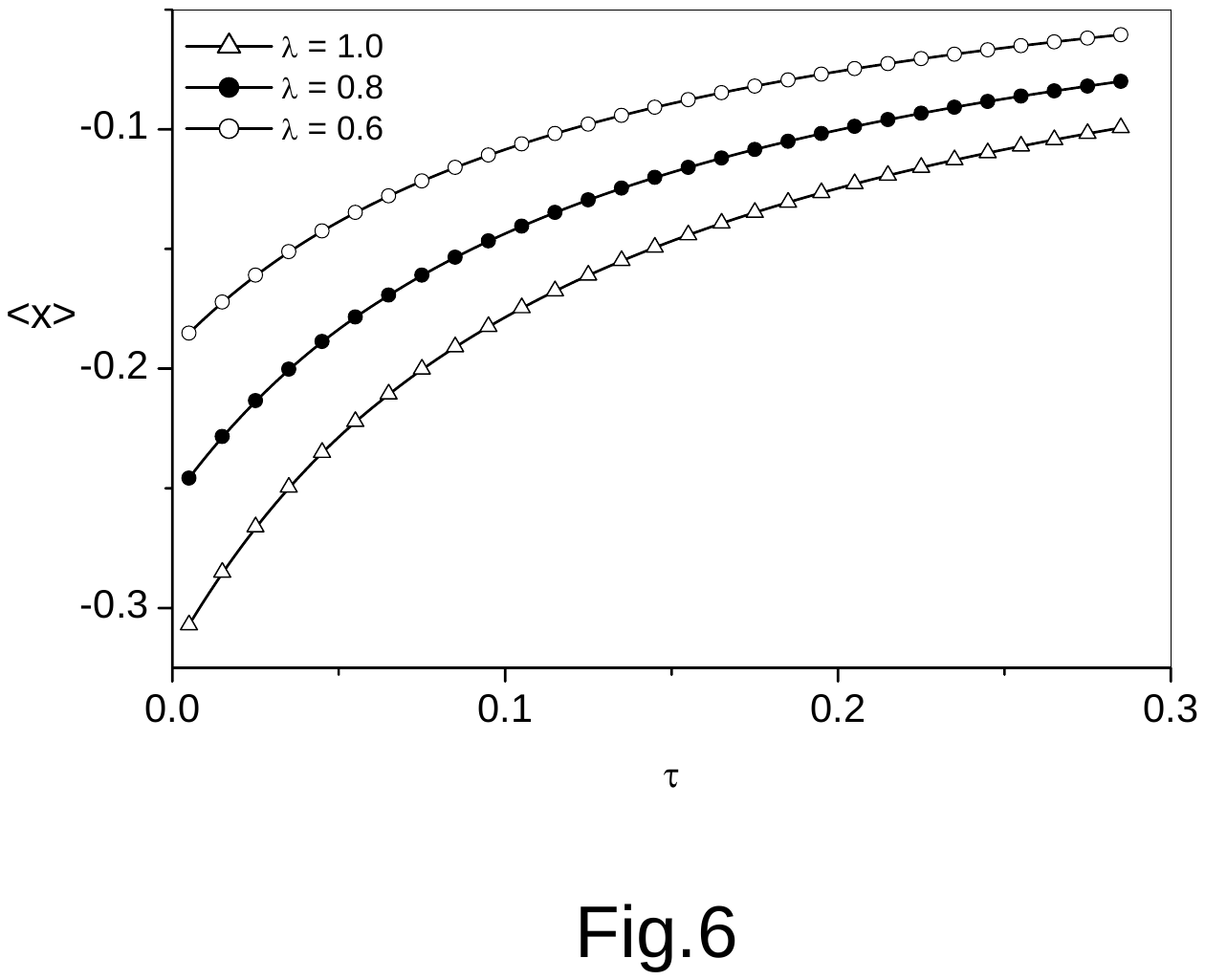}
\end{figure*}

\begin{figure*}[tp]
\centering
\includegraphics*[width=20.2cm]{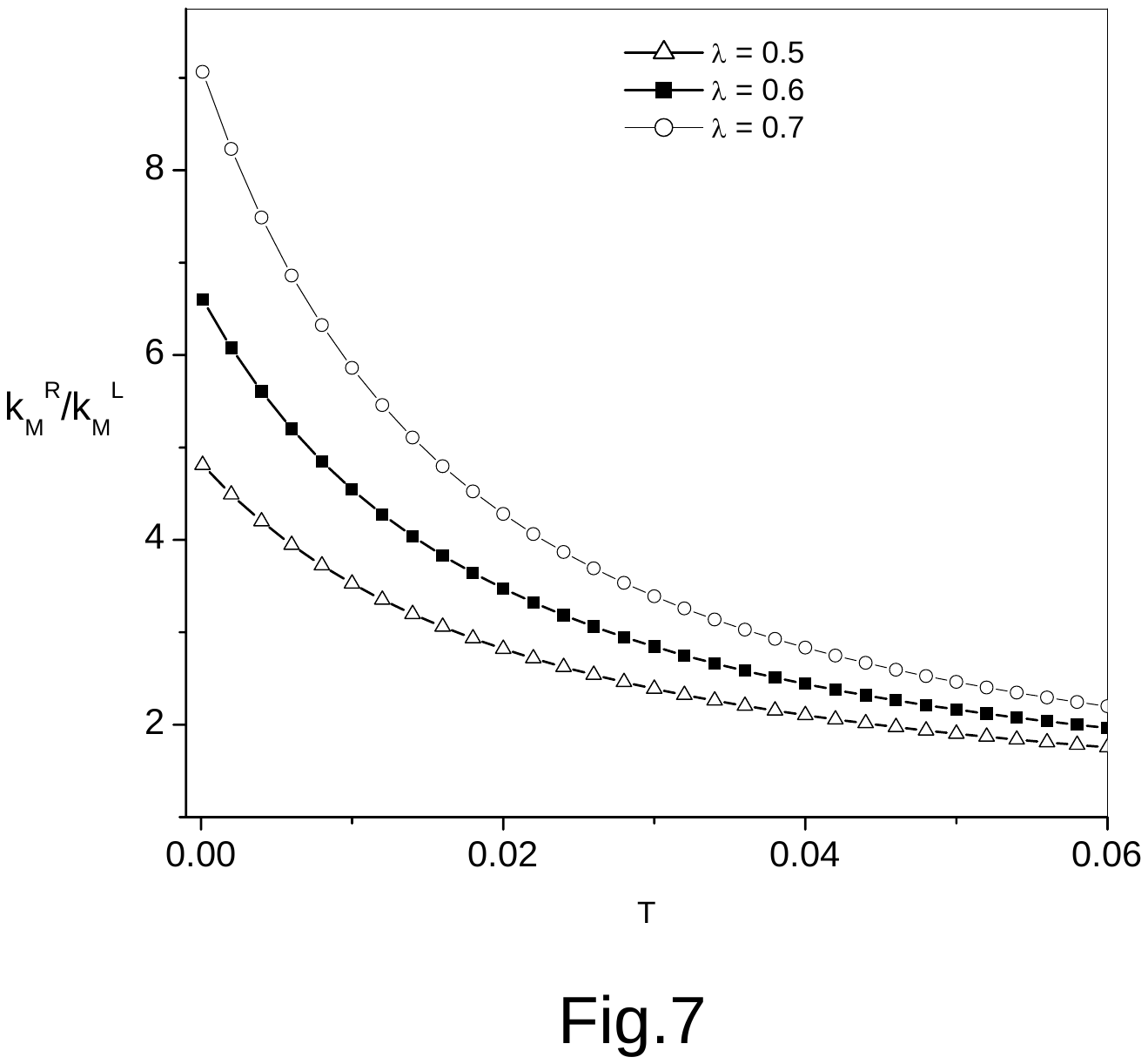}
\end{figure*}

\begin{figure*}[tp]
\centering
\includegraphics*[width=20.2cm]{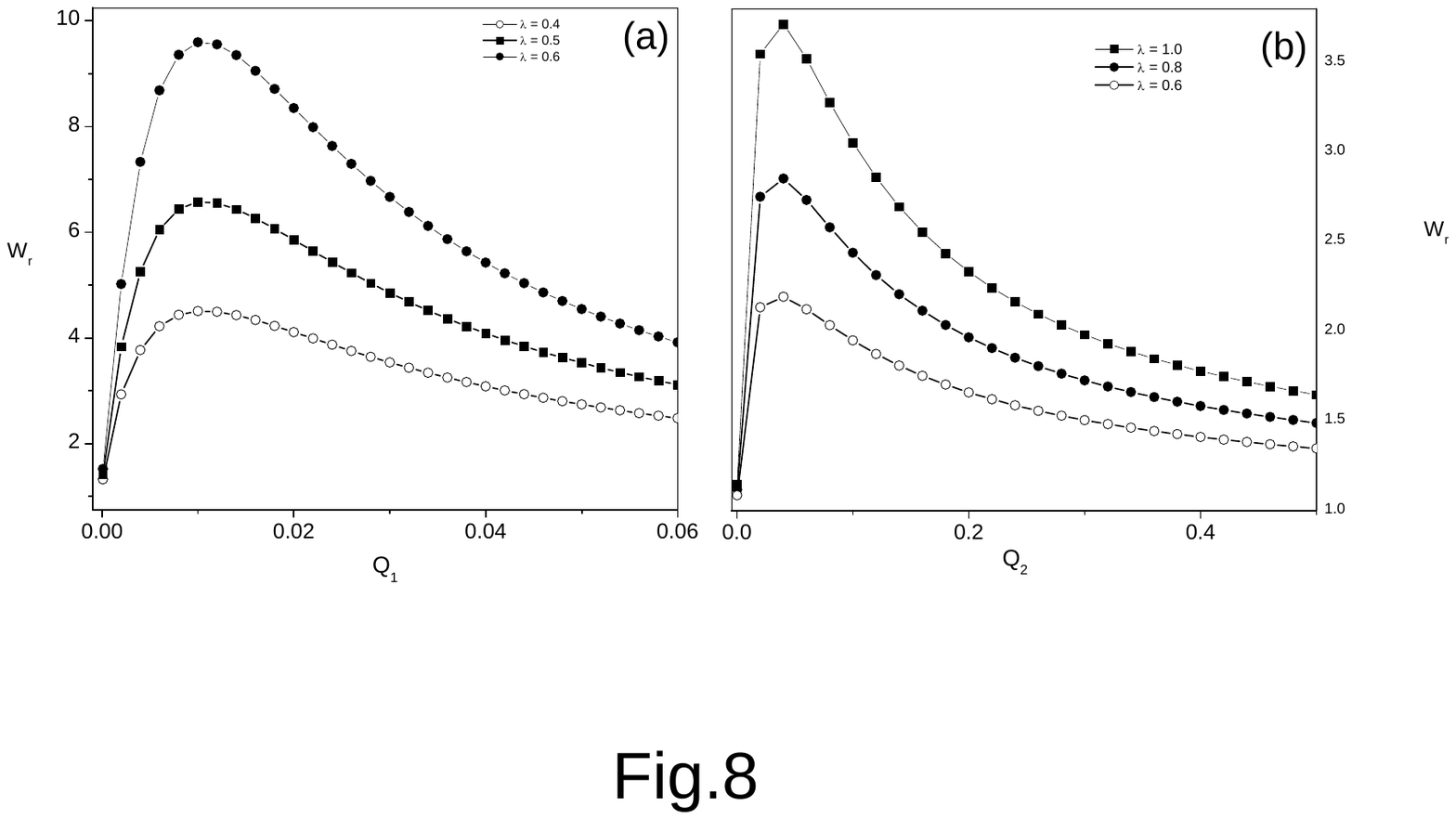}
\end{figure*}

\begin{figure*}[tp]
\centering
\includegraphics*[width=20.2cm]{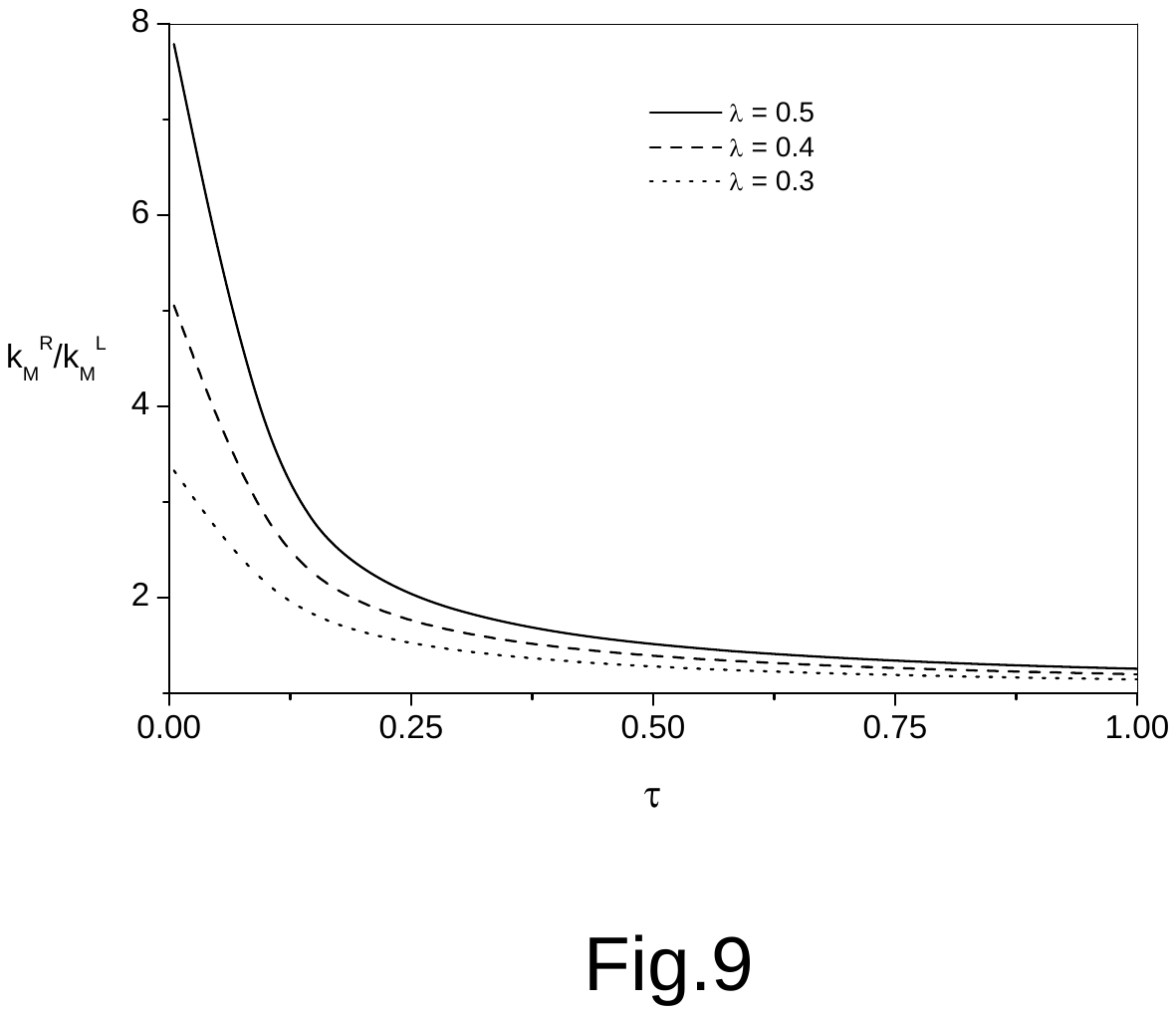}
\end{figure*}
\end{document}